\documentstyle[12pt]{article}
\renewcommand{\theequation}{\thesection.\arabic{equation}}

\newlength{\extraspace}
\newlength{\extraspaces}
\newcounter{dummy}

\newcommand{\be}{\begin{equation}
\addtolength{\abovedisplayskip}{\extraspaces}
\addtolength{\belowdisplayskip}{\extraspaces}
\addtolength{\abovedisplayshortskip}{\extraspace}
\addtolength{\belowdisplayshortskip}{\extraspace}}
\newcommand{\ee}{\end{equation}}

\newcommand{\ba}{\begin{eqnarray}
\addtolength{\abovedisplayskip}{\extraspaces}
\addtolength{\belowdisplayskip}{\extraspaces}
\addtolength{\abovedisplayshortskip}{\extraspace}
\addtolength{\belowdisplayshortskip}{\extraspace}}
\newcommand{\ea}{\end{eqnarray}}

\newcommand{\baa}{
\addtocounter{equation}{1}
\setcounter{dummy}{\value{equation}}
\setcounter{equation}{0}
\renewcommand{\theequation}{\thesection.\arabic{dummy}\alph{equation}}
\begin{eqnarray}
\addtolength{\abovedisplayskip}{\extraspaces}
\addtolength{\belowdisplayskip}{\extraspaces}
\addtolength{\abovedisplayshortskip}{\extraspace}
\addtolength{\belowdisplayshortskip}{\extraspace}}
\newcommand{\eaa}{
\end{eqnarray}
\setcounter{equation}{\value{dummy}}
\renewcommand{\theequation}{\thesection.\arabic{equation}}}

\newcommand{\ban}{\begin{eqnarray*}
\addtolength{\abovedisplayskip}{\extraspaces}
\addtolength{\belowdisplayskip}{\extraspaces}
\addtolength{\abovedisplayshortskip}{\extraspace}
\addtolength{\belowdisplayshortskip}{\extraspace}}
\newcommand{\ean}{\end{eqnarray*}}

\newcommand{\newsection}[1]{
\vspace{13mm}
\pagebreak[3]
\addtocounter{section}{1}
\setcounter{equation}{0}
\setcounter{subsection}{0}
\setcounter{footnote}{0}
\begin{flushleft}
{\sc \thesection. #1}
\end{flushleft}
\nopagebreak
\smallskip
\nopagebreak}

\newcommand{\newsubsection}[1]{
\vspace{10mm}
\pagebreak[3]

\addtocounter{subsection}{1}
\noindent{\it 
#1}
\nopagebreak
\vspace{2mm}
\nopagebreak}

\newcommand{\hf}{{\textstyle{1\over 2}}}

\newcommand{\is}{\! & \! = \! & \!}

\newcommand{\nonu}{\nonumber \\[1.5mm]}

\newcommand{\ra}{\rightarrow}

\newcommand{\ab}{{\alpha\beta}}
\newcommand{\half}{{\textstyle{1\over 2}}}

\newcommand{\X}{{\mbox{\footnotesize $X$}}}
\newcommand{\Xt}{\tilde{\mbox{\footnotesize $X$}}}

\begin{document}
\addtolength{\baselineskip}{.7mm}

\thispagestyle{empty}

\begin{flushright}
{\sc PUPT}-1380\\
{\sc IASSNS-HEP}-93/8\\
February 1993
\end{flushright}
\vspace{.3cm}

\begin{center}
{\large\sc{A Unitary $S$-matrix for 2D Black Hole\\[4mm]
Formation and Evaporation}}\\[8mm]

{\sc Erik Verlinde}
\\[3mm]
{\it School of Natural Sciences\\[2mm]
Institute for Advanced Study\\[2mm]
Princeton, NJ 08540} \\[.4cm]
{ and}\\[.4cm]
{\sc  Herman Verlinde}\\[3mm]
{\it Joseph Henry Laboratories\\[2mm]
 Princeton University\\[2mm]
Princeton, NJ 08544}
\\[15mm]

{\sc Abstract}
\end{center}

\noindent
We study the black hole information paradox in the context of a
two-dimensional toy model given by dilaton gravity coupled to
$N$ massless scalar fields. After making the model well-defined
by imposing reflecting boundary conditions at a critical value
of the dilaton field, we quantize the theory and derive the
quantum $S$-matrix for the case that $N$=$24$. This $S$-matrix is
unitary by construction, and we further argue that in the semiclassical
regime it describes the formation and subsequent Hawking evaporation
of two-dimensional black holes.
Finally, we note an interesting correspondence between the dilaton
gravity $S$-matrix and that of the $c=1$ matrix model.

\vfill

\newpage

\newsection{Introduction}

The discovery that black holes can evaporate by emitting thermal
radiation has led to a longstanding controversy about whether
or not quantum coherence can be maintained in this process.
Hawking's original calculation \cite{hawking} suggests that an
initial state, describing matter collapsing into a black hole,
will eventually evolve into a mixed state describing the thermal
radiation emitted by the black hole. The quantum physics of black
holes thus seems inherently unpredictable. However, this is clearly
an unsatisfactory conclusion, and several attempts have been made
to find a description of black hole evaporation in accordance with the
rules of quantum mechanics \cite{page,thooft2}, but so far all these attempts
have run into serious difficulties.  Classically, the loss of information
takes place at the singularity of the black hole, which forms a space-like
boundary of space-time at which the evolution stops. While it is conceivable
that quantum effects may alter this picture by smoothing out the singularity,
it still is difficult to see how the information can be retrieved from behind
the black hole horizon without a macroscopic violation of causality.
Nevertheless, it seems worthwhile to investigate this possibility.

We would like to study some of these issues with the help of a toy model,
with the hope that it will capture some of the essential features of the
full theory.  Recently, there has been considerable interest in
1+1 dimensional dilaton gravity, described by the Lagrangian \cite{cghs}
\be
\label{dilgr}
S_ = {1\over 2\pi} \int \sqrt{-g} \Bigl[e^{-2\phi}(R +
4 (\nabla \phi)^2 + 4 \lambda^2)
- \half \sum_{i=1}^N  (\nabla f_i)^2 \Bigr]
\ee
Here $\phi$ is the dilaton and $f_i$ are massless scalar fields.
The main virtue of this model is that it is completely soluble,
at least classically, while it also possesses classical
black hole solutions \cite{2dbh,withole}.
A further motivation for studying this model is
that, up to a factor of 2
in front of the dilaton kinetic term, the gravitational part
of the action (\ref{dilgr}) is identical to the spherically symmetric
reduction of the 3+1 dimensional Einstein action.

The information paradox also arises in the context of this toy model,
since on the one hand there appears to be no obvious reason why the
Lagrangian (\ref{dilgr}) would not describe a well-defined
1+1 dimensional quantum field theory. It even looks exactly soluble
and hence it should have some well-defined unitary $S$-matrix.
On the other hand, semi-classical studies \cite{cghs}--\cite{gs}
show that the 1+1 dimensional black holes in this model are unstable and
evaporate via the Hawking process. Thus dilaton gravity provides a
simple and interesting testing ground in which one can explicitly address
the paradox and decide which of the two scenarios is realized.

In this paper we will provide evidence in favor of the first possibility.
Namely, we will explicitly construct the quantum theory
described by (\ref{dilgr}) in the special case that $N\!=\!24$, and
after imposing appropriate boundary conditions, derive a
well-defined unitary scattering matrix. To explain our approach, we begin
in section 2 with a summary of the (semi-)classical properties of the model.
The quantization of dilaton gravity
 and the construction of the $S$-matrix are explained in sections
3 and 4. Finally, in section 5 we discuss some of the
properties of the $S$-matrix and clarify its physical interpretation.
We will argue that in the semiclassical limit it reproduces
the expected semiclassical physics of Hawking radiation.
We also point out an interesting correspondence with the scattering equations
of the $c=1$ matrix model \cite{c=1,polch}. Finally, we compare our
result with the black hole $S$-matrix proposed by 't Hooft
\cite{thooft2}.

\newsection{The (Semi-)Classical Model.}

Dilaton gravity has no
local gravitational degrees of freedom: the classical equations of motion
of (\ref{dilgr})
uniquely determine the metric $ds^2 = g_\ab dx^\alpha dx^\beta$ and
dilaton field $\phi$ for a given energy-momentum distribution.
Explicitly, one can choose a preferred set of
coordinates $x^\pm$ such that the classical
metric and dilaton are related via
\be
\label{met}
\qquad ds^2 = e^{2\rho} dx^+ dx^- ,\qquad \qquad
\rho= \phi.
\ee
In these coordinates the remaining equations of motion read
\ba
\label{cl}
\partial_+\partial_- e^{-2\phi} \is -\lambda^2 \nonu
\partial^2_\pm e^{-2\phi} \is T_{\pm\pm}
\ea
where $T_{\pm\pm} = \sum_i \half (\partial_\pm f_i)^2$
is the traceless matter energy-momentum tensor. These equations can
be integrated to
\be
\label{sol}
e^{-2\phi} = M -
\lambda^2 x^+ x^- - \int_{{}_{\! x^+}}^{{}^{\infty }}\!\!\!
dy^+\int_{{}_{\! y^+}}^{{}^{\infty}} \!\!\! dz^+ T_{++}
- \int^{{}^{ x^-}}_{{}_{\! -\infty}} \!\!\!
dy^-\int^{{}^{ y^-}}_{{}_{\!-\infty}} \!\!\! dz^- T_{--}
\ee
If we put the $T_{\pm\pm}=0$ this reduces to the static black hole solution
of mass $M$. The terms involving $T_{\pm\pm}$
represent the classical back reaction of the metric
due to the incoming or outgoing matter.

{\renewcommand{\footnotesize}{\small}

We are interested in the situation where a black hole is formed by
incoming matter, but where initially there is no black hole. In this case
the geometry in the far past is that of the linear dilaton vacuum
\be
\label{ldv}
\qquad \qquad ds^2 = -
{dx^+ dx^-\over \lambda^2 x^+x^-}\, ;\qquad \qquad e^{-2\phi} =
- \lambda^2 x^+ x^-
\ee
Here the coordinates $x^\pm$
are restricted to a half-line, $x^+\! >\! 0$  and $x^- \! <\! 0$;
they are related to the standard flat coordinates $r$ and $t$ for
which $ds^2 = -dt^2 + dr^2$ via
\be
x^\pm \! =\! \pm\exp(\lambda({t\pm r}))
\ee
The field $e^{\phi}$ is known to play the role of the coupling
constant of the model, and becomes infinite for $r \!\ra\! -\infty$.
It is therefore not appropriate to treat this strong coupling regime
as an asymptotic region of space-time, but rather one would need to specify
some physically reasonable boundary condition, that makes the model
well-defined.\footnote{Note that in the spherically symmetric reduction of
the Einstein theory, the line $e^{-2\phi}=0$ coincides with the origin
at $r=0$ and indeed defines a reflecting boundary.}

Before we can formulate these boundary conditions, we first
need to recall some properties of the semiclassical theory.
Namely, it turns out (see \cite{rst} and section 4) that
in the quantum theory the vacuum carries a negative Casimir
energy proportional to the number of scalar fields\footnote{In
\cite{rst} this vacuum energy was taken to be proportional to $N$-$24$.
It will be shown later that (\ref{casimir}) is the correct value, at
least for $N=24$.}}
\be
\label{casimir}
\langle 0| T_{\pm\pm}|0\rangle = - {N\over 24{x^\pm}^2}.
\ee
The origin of this vacuum energy is that $T_{\pm\pm}$ in equation
(\ref{sol}) is normal ordered with respect to the `Kruskal' coordinates
$x^\pm$, while the asymptotic vacuum is defined in terms of the physical
asymptotic coordinates $r$ and $t$.
As a consequence, the vacuum solution for the dilaton
gravity fields is no longer exactly described by the linear dilaton
vacuum (\ref{ldv}). It is possible, however, to restore this property
of the classical theory by introducing a semiclassical dilaton field
$\tilde{\phi}$ and metric $\tilde{\rho}$ via
\ba
\label{redef}
e^{-2\phi} \is e^{-2\tilde{\phi}} + \textstyle{N\over 12}
\tilde{\phi}\\[2.5mm]
e^{2\rho-2\phi} \is e^{2\tilde{\rho} - 2\tilde{\phi}} \nonumber
\ea
Via this field redefinition one effectively incorporates
semiclassical corrections to the equation of motion (\ref{cl}), such as
the back reaction of the metric due to Hawking radiation \cite{cghs,rst}.
In general, however, invariant
physical quantities such as an $S$-matrix will not depend on such field
redefinitions, and we
will therefore continue to work with the original field variables
$\phi$ and $\rho$.

The boundary condition we will impose is the same one as
proposed by Russo, Susskind and Thorlacius in \cite{rst},
and is motivated by the following observation. The right-hand
side of (\ref{redef}) attains a minimum when $e^{-2\tilde{\phi}}
= {N\over 24}$, and thus only values of $\phi$ larger than a
certain critical value $\phi_{cr}$ have a semiclassical interpretation.
In other words, the semiclassical theory becomes singular on a certain
critical line $x^\pm=\X^\pm(\tau)$ at which
\be
\label{bond}
\phi(\X^+,\X^-) = \phi_{cr}.
\ee
This critical line is timelike for sufficiently small perturbations around
the vacuum, and it
is thus natural to identify it  with a physical
boundary of the 1+1-dimensional space-time at which we can impose
reflecting boundary conditions for the $f$-fields.
There are many possible choices, but we will for definiteness take the
simplest choice and impose Neumann boundary conditions.
This choice is coordinate invariant, and implies that there is no
net matter energy-momentum flux through the critical line
\be
\label{reflex}
T_{--} (d\X^-)^2
= T_{++}(d\X^+)^2.
\ee
A natural set of corresponding boundary conditions on the gravitational
sector, that are also coordinate invariant, is to demand that the derivative
of the dilaton must vanish both along and perpendicular to the boundary
\be
\label{RST}
\partial_\pm e^{-2\phi}(\X^+,\X^-) = 0.
\ee
The authors of \cite{rst} arrived at an identical set of boundary
conditions by imposing the requirement that the
semiclassical metric, obtained via (\ref{redef})
remains nonsingular at the boundary, and therefore
interpreted these conditions as the implementation of cosmic censorship
at the critical line $\phi$=$\phi_{cr}$.

As long as the boundary is time-like, its trajectory is completely
determined by the incoming energy flux.
By combining equations (\ref{sol}) and (\ref{RST}), one finds
that the boundary trajectory is given by the following
elegant equations
\ba
\label{nonu}
\lambda^2\X^- \is + \int_{{}_{\! X^+}}^{{}^{\infty}} \!\!\!
dx^+\, T_{++}, \\[2.5mm]
\lambda^2\X^+ \is - \int^{{}^{X^-}}_{{}_{\!-\infty}} \!\!\!\!
dx^- \,T_{--}.\nonumber
\ea
Here the second relation follows from the first one by
the reflection condition (\ref{reflex}).

The equations (\ref{nonu}) will play a central
role in the following.
As an aside, let us point out that they
can in fact be derived from a very
simple action principle as follows. Since the boundary
trajectory $\X^\pm(\tau)$ represents the only dilaton gravity degree of
freedom that couples directly to the matter fields,
it should be allowed to eliminate all other gravitational fields
from the action via gauge invariance or their equation
of motion. If we follow this procedure, we find
that the gravitational action can be reduced to a pure boundary term
\be
\label{sb}
S_b =
\lambda^2\int\! d\tau \,\X^+\dot{\X}^-
\ee
which defines the free dynamics of the critical line.
The coupling to matter is described simply by the restricting
the integral over $x^\pm$ in the matter action to the right of
the boundary. The equations (\ref{nonu}) for $\X^\pm$
are then obtained by writing the variation of the
matter action as
\be
\delta S_m = \int\! d\tau\,
\Bigl(T_{++} {\dot{\X}^+}\delta \X^+ +T_{--}
{\dot{\X}^-} \delta \X^-\Bigr)
\ee
and integrating the resulting equation of motion $\delta S_b + \delta S_m = 0$
once with respect to $\tau$. Thus the model can in a way be thought of as
a single quantum mechanical mirror interacting with a free field theory.

When we include the vacuum contribution (\ref{casimir}) in
(\ref{nonu}) we see that semiclassically there is a low-energy regime
for which the boundary stays time-like. In this
regime, therefore, there is a well-defined scattering problem
that determines the outgoing matter waves from a certain given
configuration of incoming matter.
However, as soon as the incoming energy density exceeds
the Casimir energy of the vacuum, the
above semiclassical equations break down and the critical line
goes over in a space-like black hole singularity.

This fact appears
to be a serious problem in defining the quantum theory, since
it is not meaningful to impose reflection conditions on a space-like
boundary. Thus at high energies the theory
still seems to have the usual problem of information loss.
However, near the black-hole singularity quantum fluctuations will be
very large and it is not known what new physics may occur here.
Hence, to define the model in this regime, we are free to make some
assumptions, as long as we do not (drastically) change the known
laws of physics in the semiclassical region of space-time.
In particular, because the metric loses its classical
meaning near the singularity, we are allowed to assume that in the
quantum theory the boundary actually remains time-like.
In the following section we work out this idea to construct
an $S$-matrix for quantum dilaton gravity.

\newsection{Quantum Dilaton Gravity.}

We will now investigate how this semiclassical situation translates to
the quantum level. The strategy will be to set up the quantum
theory in the low energy regime, where we will adopt the above
boundary condition to make the model well-defined.
Our aim is to find the quantum mechanical $S$-matrix that describes the
scattering of matter off the dynamical boundary.
To avoid unnecessary complications due to the conformal anomaly we
will restrict our discussion to the critical case, which is
dilaton-gravity with 24 scalar fields $f_i$. This special model is by far
the simplest, while it still contains all the essential physics.
In particular, as will be explained later, it has a semiclassical
regime in which black holes are formed and evaporate by emitting
Hawking radiation, even though the total conformal anomaly cancels.

We will first describe the quantization of pure dilaton-gravity in the
conformal gauge. Later we will combine it with the
matter fields (and ghosts) to determine the physical spectrum.
In the conformal gauge
\be
\label{confg}
ds^2 = e^{2\rho} du\,dv
\ee
the action of the pure dilaton-gravity theory becomes
\be
\label{Sconf}
S={1\over \pi}\int \!dudv\, e^{-2\phi}
(2\partial_u\partial_v\rho-4\partial_u\phi\partial_v\phi + \lambda^2
e^{2\rho}) ,
\ee
and the equations of motion can be written as
\ba
\label{cons}
\partial_u\partial_v(\rho-\phi)  \is 0  \\[2mm]
\partial_u \partial_v e^{-2\phi} \is \lambda^2 e^{2\rho-2\phi}. \label{bla}
\ea
The action (\ref{Sconf}) defines for all values of $\lambda^2$
a conformally invariant field theory. In the quantum theory
the modes of the  energy-momentum tensor
\be
\label{Tg}
T^g_{uu} = (4\partial_u \rho\partial_u \phi-2\partial_u^2\phi)e^{-2\phi}
\ee
generate a Virasoro algebra with central charge $c\!=\!2$.
Furthermore we know that the operator $e^{2\rho-2\phi}$ representing the
cosmological constant must be marginal.

We consider the theory on the Minkowski half-plane
with a {\it fixed} time-like boundary given by $u\!=\!v$.
The coordinates $u$ and $v$ differ from the coordinates $x^\pm$ of
the previous section by a conformal transformation, that
depends on conformal factor $\rho$ and
the dilaton field $\phi$.
We will now show that in the
quantum theory the $x^\pm$-coordinates will appear as a pair of
free scalar fields $\X^\pm$, and that
{\renewcommand{\footnotesize}{\small} that the complete
dilaton gravity can be conveniently
reformulated in terms of these fields.
\footnote{\renewcommand{\footnotesize}{\small}The following construction
in fact follows naturally from the gauge theory formulation of dilaton gravity
described in \cite{hv}.}}

The first equation of motion (\ref{cons}) implies that the marginal
operator $e^{2\rho -2\phi}$ factorizes as a product of two chiral components.
Therefore, since it has conformal dimension (1,1), we can introduce
two chiral scalar fields $\X^\pm$  and write
\be
e^{2\rho-2\phi}=\partial_u\X^+\partial_v\Xt^-. \label{rhoXX}
\ee
Classically the chiral fields $\X^+(u)$ and $\Xt^-(v)$ indeed represent
the conformal transformation that maps $(u,v)$ on
to the coordinates $(x^+,x^-)$ in which $\rho\!=\!\phi$.
Next we insert this into the second equation of motion (\ref{bla})
and the solution after integrating once can be written as
\ba
\label{phiXX}
\partial_u e^{-2\phi} \is \lambda^2\partial_u \X^+(\X^-\! -\Xt^-)\nonu
\partial_v e^{-2\phi} \is -\lambda^2(\X^+\!-\Xt^+) \partial_v \Xt^-.
\ea
Here we introduced two additional chiral fields $\X^-(u)$ and
$\Xt^+(v)$, that naturally can be combined with $\X^+$ and $\Xt^-$
to obtain  two ordinary free scalar fields.
The boundary conditions
on the dilaton gravity fields become
very simple in the new variables: the condition that
$\partial_u{e^{-2\phi}}=\partial_v{e^{-2\phi}}=0$
at $u=v$ simply translates in to
\be
\X^\pm(u)_{\strut{|u=v}}=\Xt^\pm(v)_{\strut{|u=v}}.
\ee
Since this boundary condition identifies the left- and  right-movers we may
from now on drop the tildes and work only with, say, the
left-movers $\X^\pm(u)$.

Equations (\ref{rhoXX}) and (\ref{phiXX}) are not just valid as
classical field redefinitions, but with a suitable normal ordering
prescription they represent well-defined quantum identifications of
operators. This means that the correlation functions
and operator algebra of the dilaton field and the conformal factor are
in principle determined by those of $\X^\pm$.
The operator algebra
of the fields $\X^\pm$ is most easily obtained by noting that, after
substituting the redefinitions into (\ref{Tg}), the gravitational
energy-momentum tensor is given  by the
familiar free field expression
\be
\label{TXX}
T^g_{uu}= \lambda^2 \partial_u \X^+\partial_u \X^-.
\ee
{}From the fact that  the fields $\X^\pm$ must have the usual
scalar operator algebra with the energy-momentum tensor $T^g_{uu}$,
it follows that the operators $\X^+$ and $\X^-$
satisfy the standard free field commutation relations
\be
\label{comm}
[\partial_u \X^\pm(u_1), \X^\mp(u_2)] =
\lambda^{-2} \delta(u_{12}),
\ee
with $u_{12} = u_1-u_2$.
Thus we have indeed mapped the pure dilaton gravity theory onto a
theory of two free scalar fields. Another
method to derive this result is to compute the action
$S(\X^\pm)$ by substituting (\ref{rhoXX}) and  (\ref{phiXX}) into
(\ref{Sconf}); one obtains the standard free scalar field action.

The correspondence with the semiclassical discussion of the previous
section requires that the fields $\X^\pm$ are asymptotically identified
with the coordinates $x^\pm$  of the linear dilaton vacuum (\ref{ldv}).
This tells us that when $u$ runs from $-\!\infty$ and
to $\infty$, $\X^+(u)$ should go from $0$ to $\infty$ while $\X^-(u)$
must go from $-\infty$ to $0$ and further that each behaves
asymptotically as $e^{\lambda u}$, resp. $e^{-\lambda u}$. Because of these
somewhat unconventional asymptotic conditions
we can not simply use the standard mode-expansion for the scalar fields
$\X^\pm$ to construct the dilaton gravity Hilbert space. We find that the only
mode-expansion that is consistent with the required asymptotic
behaviour is of the form
\newcommand{\Xmode}{\mbox{\sl x}}
\be
\label{xpans}
\partial_u \X^\pm(u)= {e^{\pm \lambda u} \over {\sqrt 2}} \ +
{e^{\pm \lambda u} }\int d\omega \, \Xmode^\pm(\omega)\, e^{-i\omega u},
\ee
where the modes $\Xmode^\pm(\omega)$ satisfy
\be
[\Xmode^+(\omega_1),\Xmode^-(\omega_2)] =  \lambda^{-2}(\omega + i\lambda)
\delta(\omega_1 \! + \! \omega_2).
\ee
This leads to the following Green function
\be
\label{greent}
\langle 0 |  \X^+(u_1) \X^-(u_2)|0 \rangle
=
\lambda^{-2} \int_{{}_{\!- \infty}}^{{}^{\lambda u_{12}}}\!
dx\,{ e^{x}\over x}
(1 -\hf x),
\ee
where $|0\rangle$ denotes the vacuum state that is annihilated by
all modes $\Xmode^\pm(\omega)$ with $\omega\! >\! 0$.
This Green function has the right asymptotic behaviour for large $u_{12}$,
while its behaviour at short relative distances has been adjusted such that
the conformally normal ordered energy-momentum tensor (\ref{TXX}) has no
vacuum expectation value. This last requirement
ensures that the state $|0\rangle$ describes the physical vacuum
and  fixes the coefficient in front of the
first term in (\ref{xpans}).

In principle, all correlation functions of the original field
variables $\rho$ and $\phi$ can now be obtained from the free field
correlators of $\X^\pm$ via the identifications (\ref{rhoXX}) and
(\ref{phiXX}). However, some special care is required in regularizing
these composite operators. For example, to ensure that the right-hand side
of (\ref{phiXX}) correctly behaves as a dimension 1 conformal field,
we need to define it as
\be
\label{reg}
\X^-\partial_u \X^+
=  :\! \X^- \partial_u\X^+\! : -
{\textstyle{1\over 2\lambda^2}} \partial_u \log \partial_u \X^+
\ee
where $: \ :$ denotes usual normal ordening.
One should keep in mind, however, that only the expectation values
of conformally invariant operators have a precise physical meaning.

\newsection{The S-matrix and the Light-Cone Gauge.}

Let us now include the matter fields $f_i$.  Similar as for the gravitational
fields, the reflecting boundary condition at $u=v$ gives an identification
between the left and right moving parts of the fields $f_i$, so we may
again work with just the left-movers $f_i(u)$. Because we are working in
the conformal gauge we have the usual condition
that the sum of the matter and gravitational energy-momentum tensor
vanishes. This implies
\be
\label{constraint}
\lambda^2 \partial_u\X^+\partial_u\X^- = T^m_{uu}.
\ee
We could impose this condition on physical states
in the form of Virasoro constraints, or equivalently, introduce ghosts
and demand that physical states and operators are BRST-invariant.
Only for the critical theory with 24
scalar fields $f_i$ the BRST-charge $Q$
is nilpotent without the need of adding a one-loop correction
to the gravitational energy-momentum tensor.

The reader may have noticed that our formulation of quantum dilaton gravity
theory is very similar to critical open string theory, with the
matter fields playing the role of the transverse string coordinates while
the $\X^\pm$ are like the light-cone string coordinates. As we will see
momentarily, this correspondence with open string theory proves to be
very useful in constructing the $S$-matrix.

A convenient way to describe the physical Hilbert space is to choose the
analogue of the light-cone gauge and use the residual conformal symmetry to
introduce a physical time coordinate that is defined in terms of either
$\X^+$ or $\X^-$. In fact, for our purpose the light-cone gauge is
more than just a convenient choice, but has a precise physical significance:
it can be seen that a past observer will identify as the
proper time-coordinate along past null infinity the variable
\be
\label{lcg}
\tau_+ = \lambda^{-1}\log(\X^+),
\ee
while a future observer will identify
\be
\label{lcgb}
\tau_- = - \lambda^{-1}\log(-\X^-)
\ee
as the proper time-coordinate along future null infinity. These two choices
each define an allowed light-cone gauge condition, and each lead to a
different description of the same physical Hilbert space.
The past observer will use the physical coordinate (\ref{lcg}) to
define creation- and annihilation operators by decomposing the $f_i$ fields
in modes as
\be
\label{expans}
f^\prime_i(\tau_+) = \int \! {d\omega\over 2\pi}
\, \alpha_i(\omega)\, e^{i\omega\tau_+},
\ee
with
$[\alpha_i(\omega_1),\alpha_j(\omega_2)] =
\omega_1\delta_{ij} \delta(\omega_{12})$.
The in-vacuum is annihilated by all $\alpha_i(\omega)$ with $\omega\!>\!0$,
while the $\alpha_i(\omega)$ with $\omega \!<\! 0$
create the incoming particles.
The resulting states are all physical.
On the other hand,
a future observer, who detects the outgoing particles, will use the physical
coordinate (\ref{lcgb}) to write
\be
\label{expansb}
f^\prime_i(\tau_-) = \int \! {d\omega\over 2\pi}
\,\beta_i(\omega) e^{i\omega\tau_-},
\ee
with $[\beta_i(\omega_1),\beta_j(\omega_2)] = \omega_1\delta_{ij}
\delta(\omega_{12})$, and use these modes to construct the natural
out-basis of physical states.
Both constructions are the direct analogue of the standard
light-cone description of the physical Hilbert space of the
open string. In the covariant formalism, $\alpha_i(\omega)$
and $\beta_i(\omega)$ define conformally invariant vertex operators,
given by
\ba
\label{ain}
\alpha_i(\omega) \is
\int\! du \,   f^\prime_i(u)\, ( \X^+)^{i\omega/\lambda} \\[2.5mm]
\label{aout}
\beta_i(\omega) \is
\int\! du \,   f^\prime_i(u)\, (-\X^-)^{-i\omega/\lambda}
\ea
They can be compared with the DDF operators \cite{DDF}, which are known
to generate the complete physical spectrum. Based on this analogy,
it seems a reasonable assumption that also in our case the $\alpha$ and
$\beta$ oscillators
each separately constitute a complete basis of physical operators.

Thus we now arrive at a very simple characterization of the
scattering matrix of dilaton gravity. Namely, it
is nothing other than the unitary transformation that interchanges the
role of $\X^+$ and $\X^-$ and maps the first
light-cone basis of physical states to the second basis. In other words,
$S$ is the intertwiner between the $\alpha_i$ and $\beta_i$ oscillators
\be
\label{intertwine}
\alpha_i(\omega) \ S  \, = \,  S \ \beta_i(\omega) ,
\ee
which, being a canonical transformation, is guaranteed to define
a unitary operator. Note that this $S$-matrix commutes with the energy
operator,
so in particular it maps the in-vacuum onto the out-vacuum.
Matrix elements of $S$
\be
\label{melt}
\langle in | out \rangle \, = \, \langle 0 | \prod_k \alpha_{i_k}(\omega_k) \,
\prod_l \beta_{i_l}(\omega_l) | 0\rangle
\ee
can thus be written as integrated correlation functions of the covariant
vertex operators (\ref{ain})-({\ref{aout}) in the free field theory
defined by the $f$ and $\X^\pm$-fields \cite{toappear}.

To make the relation between the two types of modes more explicit,
we can write the formula (\ref{aout}) for the outmode $\beta_i$ in the
light-cone gauge $u=\tau_+$ and solve for $\X^-(\tau_+)$ by using the physical
state condition (\ref{constraint}). In this way we can express
the right-hand side in terms of the in-modes $\alpha_i$. The exact quantum
identification of $\X^-(\tau_+)$ can be found by some standard
technology of light-cone gauge string theory \cite{brower}.
We define $\X^-(\tau_+)$ via a fourier mode expansion
\ba
\X^-(\tau_+) \is e^{-\lambda\tau_+}
\int \! {d\omega \over \lambda+i\omega} \hat{\X}^-(\omega) \, e^{i\omega\tau_+}
\label{xmint}
\\[3mm]
\hat{\X}^-(\omega) \is \lambda \int
\! du \, \partial_u \X^- (\X^+)^{1-i{\omega/\lambda}}.
\ea
where the modes $\hat{\X}^-(\omega)$ are physical operators,
provided the composite operator is regularized as in (\ref{reg}).
Physically, the operator $\X^-(\tau_+)$ represents the space-time
trajectory of the critical line (\ref{bond}).
Now, a straightforward calculation
shows \cite{brower,toappear} that the modes $\hat{\X}^-(\omega)$
satisfy the commutation relations
of a Virasoro algebra with central charge $c\! =\! 24$
\be
\lambda^2 [\hat{\X}^-(\omega_1),\hat{\X}^-(\omega_2)] =
(\omega_1\! - \! \omega_2)
\hat{\X}^-(\omega_1\! + \! \omega_2) +
2 \omega_1(\omega_1^2+\lambda^2)
\delta(\omega_1\! +\! \omega_2).
\ee
As in critical string theory, this
result is sufficient to guarantee that, within the physical Hilbert
space, we can indeed solve (\ref{constraint}) and
identify the modes $\hat{\X}^-(\omega)$
with corresponding modes of the physical energy-momentum tensor
\ba
\label{xminus}
\lambda^2
\hat{\X}^-(\omega) \is  L^{in}(\omega) - \lambda^2 \delta(\omega) \\[3mm]
L^{in}(\omega) \is \half
\int \! d\xi \, : \alpha_i(\xi)\alpha_i(\omega-\xi):  \nonumber.
\ea
The term $\lambda^2\delta(\omega)$ in (\ref{xminus}) is
needed to ensure that the central term in the algebra on both sides
has the same form. It represents the constant vacuum contribution
(\ref{casimir}) to the energy density.

Inserting the identification (\ref{xminus}) into (\ref{xmint})
and (\ref{aout}) in principle gives an expression for
the outgoing modes $\beta_i(\omega)$ in terms of the
ingoing modes.\footnote{The normal ordening prescription for this
expression is uniquely fixed by the covariant definition of the $S$-matrix.}
This procedure is the full quantum
version of the scattering off of the dynamical boundary
$\phi$ = $\phi_{cr}$, described in section 2.
Indeed, the above solution (\ref{xminus}) for $\X^-$
can be recognized as the mode expansion of the boundary trajectory
(\ref{nonu}), with the vacuum contribution included.

\newsection{Discussion.}

In the previous section we outlined a method for obtaining
a scattering matrix for quantum dilaton gravity.
The construction works for $N=24$ scalar fields, but we see no fundamental
difficulty to generalize it to other values of $N$. In the low energy regime,
in which the incoming energy flux stays below the critical value,
the $S$-matrix  has a clear unambiguous semiclassical interpretation as
describing the scattering of $f$-fields off the critical line (\ref{nonu}).
In terms of the proper asymptotic coordinates $\tau_\pm$,  this scattering
equation relates the ingoing and outgoing matter modes via
\ba
\label{boem}
f_i^{in}(\tau_+) \is f^{out}_i(\tau_-) \\[3mm]
\label{taupm}
\tau_- - \tau_+  \is -  \lambda^{-1}\log[1 - P_\pm(\tau_\pm)] ,
\ea
where we introduced the notation
\ba
\label{PP}
P_+(\tau_+) \is {\kappa} \int_{\tau_+}^\infty \!\!
d\sigma \, e^{\lambda(\tau_+-\sigma)}\ T^{in}_{\sigma\sigma}
\nonu
P_-(\tau_-) \is  {\kappa} \int_{-\infty}^{\tau_-}\!\!
d\sigma \, e^{\lambda(\sigma-\tau_-)}\ T^{out}_{\sigma\sigma} ,
\ea
with $\kappa = {24\over N \lambda }$.
In the above equations (\ref{PP}) the energy-momentum tensors
$T^{in}$  and $T^{out}$ are taken to be normal ordered
with respect to the physical vacuum, so the Casimir energy (\ref{casimir})
is explicitly present  in (\ref{taupm}).
In the low energy regime
\be
\label{low}
T^{in}_{\tau\tau}
< {N\lambda^2\over 24}
\ee
the boundary line  remains time-like and
 the relation (\ref{taupm}) between $\tau_+$ and $\tau_-$
is a single-valued diffeomorphism of the real line.

The interesting regime of dilaton gravity,  where black
hole formation and evaporation is expected to take place, is however
at high energies. Indeed, when the energy flux exceeds the bound
(\ref{low}) the relation (\ref{taupm}) between $\tau_+$  and $\tau_-$
is no longer single-valued,  and, as discussed in section 2, this
degeneration can be interpreted as the formation of a black hole.
In this case the semi-classical scattering equations (\ref{taupm})
do not give an invertible map from the ingoing to out-going matter waves.
This does by no means imply, however, that our quantum construction
of the $S$-matrix will not be valid in this supercritical regime.

In fact, we have given convincing arguments, based on gauge invariance,
why we have a unitary $S$-matrix that is defined on the complete physical
Hilbert space. To fully establish this fact, however, we need to show that
the physical states that we constructed indeed form a complete basis.
The close analogy with open string theory should be helpful in this
respect, since for
this case the corresponding problem was solved long ago, and is known
as the no-ghost theorem.  A technical difficulty in trying to copy
the standard no-ghost theorem appears to be that in our case the
energy spectrum is continuous, while the string spectrum is discrete.
This difficulty can however be removed simply by putting the system
in a finite box of length $L$. We plan to present further details of this
calculation  in a future publication \cite{toappear}.

\newsubsection{Relation with $C=1$ Matrix Model.}

There is a remarkable correspondence between
the present formulation of dilaton gravity and the matrix model
of two-dimensional string theory \cite{c=1}.
To explain this relation, let us rewrite the relation (\ref{nonu})
describing the scattering of the energy-momentum flux off the critical line
as follows
\be
\label{scats}
P_-(\tau) = P_+\Bigl(\tau - \lambda^{-1}\log[1-P_-(\tau)]\Bigr)
\ee
with $P_\pm$ defined as in (\ref{PP}).
The reader familiar with recent developments in two-dimensional
string theory will now recognize this equation as the classical
scattering equation of tachyons \cite{polch}.  Namely, in the matrix model,
scattering amplitudes in two-dimensional string theory are
described in terms of a free fermion field theory \cite{c=1},
in which the string tachyon modes correspond to deformations of the fermi sea.
By considering the time evolution of these perturbations,
Polchinski derived a classical scattering equation, whose form is {\it
exactly} identical to (\ref{scats}), where $P_\pm$ are identified with the
$\tau_\pm$-derivative of the in- and outgoing tachyon field \cite{polch}.
This suggests therefore an interesting interpretation of the matrix model
in which the phase space trajectory of the fermi sea plays the same role
as the critical line trajectory $\phi$ = $\phi_{cr}$ in the $\X^\pm$ plane.

In both theories, the scattering equation (\ref{scats})
defines a canonical transformation. To show this in the case
of dilaton gravity, one can use (\ref{scats}) to express the fourier
modes of the outgoing energy momentum tensor in terms of the incoming
field $P_+$ as
\be
L^{out}(\omega) = \int d\tau e^{i\omega \tau}
(1-P_+(\tau))^{1-i{\omega/\lambda}}
\ee
A straightforward calculation  \cite{toappear}
then shows that the operator on the
right-hand side indeed satisfies the Virasoro algebra,
at least at the Poisson bracket level and provided the relation between
$\tau_\pm$ is invertible.
In the matrix model, on the other hand, $P_\pm$ generate a $U(1)$ current
algebra \cite{c=1,polch}. Hence, while the
scattering equations are identical, the canonical structures are
different.  In both cases, however, the $S$-matrix is characterized as
the (unique) unitary quantum operator that implements the canonical
transformation (\ref{scats}) in the Hilbert space of the theory.
In this sense, the dilaton gravity $S$-matrix defines a
deformation of that of the $c=1$ matrix
model.{\renewcommand{\footnotesize}{\small}\footnote{\renewcommand{\footnotesize}{\small}
It may be possible,
however, to define a suitable large $N$ limit of dilaton gravity in which
the correspondence with the $c=1$ matrix model becomes exact.}}

This correspondence between the two models teaches us
some important lessons. In particular we learn that even if a classical
scattering equation exhibits pathologies of the type discussed above,
it can still lead to a well-defined unitary $S$-matrix. In the case of the
matrix model this is guaranteed via the mapping to a free fermion theory,
but also the bosonic formulation of the quantum theory does apparently
not break down even when the classical equations degenerate.
This supports our belief
that the construction of quantum dilaton gravity
described in sections 3 and 4 remains valid in the high energy regime,
where black hole formation and evaporation take place.

\newsubsection{The Correspondence Principle and Hawking Radiation}

In our formulation of quantum dilaton gravity we had to make some
assumptions about what happens in the strong coupling regions of space-time.
We should make sure, of course, that in making these assumptions we
have not inadvertently modified physics in the (semi-)classical regions in
an unacceptable way. In other words, we must verify that our quantization
procedure satisfies the correspondence principle,
in the sense that it reproduces the known semiclassical
physics of black holes. In particular, when one sends in a large energy
pulse producing a macroscopic black hole, one would like to see that
most of the outgoing matter is emitted in the form of approximately
thermal radiation.

{}From equation (\ref{taupm}) one can see that the criterion that
determines if a black hole is macroscopic is whether or not
the following condition
\be
\label{cond}
P_+(\tau_+) < 1
\ee
is violated for some value of $\tau_+$. If the energy flux exceeds
the critical value (\ref{low}) while the above condition remains
satisfied, the black hole is in general microscopic and
exists only for a rather short time. On the other hand if (\ref{cond})
is violated the black hole will be macroscopic and
exists for a very long time.

Using the analogy with the moving mirror problem \cite{birdav}, it is
now not hard to convince
oneself that the outstate will indeed contain a regime
describing Hawking radiation. If we assume that the large
incoming energy pulse is concentrated
in a small time interval, then for earlier times $\tau_+$ the integral
$P_+(\tau_+)$ in (\ref{PP}) will be of the form
\be
P_+(\tau_+) = p_+ e^{\lambda\tau_+}
\ee
where $p_+ = \kappa \int d\sigma e^{-\lambda\sigma} T^{in}_{\sigma\sigma}$.
Thus for values $\tau_+$ smaller than $-\lambda^{-1}\log p_+$,
the in- and
outgoing modes are related via a well-defined diffeomorphism
\be
\tau_- = \tau_+ - \lambda^{-1}\log(1 - p_+ e^{\lambda \tau_+})
\ee
from the
interval $\tau_+ \! < \! -\lambda^{-1}\log p_+$ to
the real line. The physical effect
of precisely this diffeomorphism was analyzed in \cite{gidnel}
and we can use their results to conclude that, in the leading semiclassical
approximation, the outgoing matter approaches a thermal spectrum for late
$\tau_-$.

This indicates that the quantum theory of sections 3 and 4 indeed
describes the right physics in the semiclassical limit.
It would of course be
more interesting to have an explicit form of the $S$-matrix
that would manifestly exhibit these features. Also,
one would like to understand better how the information that
went in the black hole is encoded in the outgoing radiation.
An important question, for example, is how long it will take for all
the information to come out.

Another interesting question is by what mechanism the information
is retrieved from behind the black hole horizon.
A closer examination of the model defined in section 3 gives
a partial answer to this last question. Namely, it can be seen
that, because the boundary is everywhere timelike in the
$(u,v)$ plane, it effectively implements the ``conveyor belt''
scenario by allowing an acausal flow of information
along the singularity. Mathematically, this can happen because the
conformal mapping $\X^-(v)$ or $\X^+(u)$ can become non-invertible
in the strong coupling region, so that signals, which always propagate
causally in the ($u,v$) plane, may propagate backwards in the physical
time defined by $\X^\pm$. Quantum mechanically, this means that the
causal structure of space-time becomes fuzzy and indeterminate near
the singularity. The key remaining problem is to show that these
strong coupling effects have no catastrophic consequences in the
semiclassical regime.

\newsubsection{Comparison with 't Hooft's Black Hole $S$-matrix.}

An important question is to what extent this approach can be
generalized to address the information paradox for four-dimensional
black holes. The main new ingredient in that case is that
the fields in addition depend on two angular coordinates.
In this connection it is interesting to point out
the similarity between our result and 't Hooft's black hole
$S$-matrix \cite{thooft2}.

The central idea of 't Hooft's proposal is that
information about the infalling particles is transferred to
the outgoing particles via high-energy collisions near the horizon.
In particular, he has shown that once the gravitational back-reaction
is taken into account, the black hole horizon becomes a
dynamical fluctuating surface, described by two light-cone
coordinate fields $\X^+(u,\theta,\phi)$ and $\X^-(v,\theta,\phi)$.
The interaction of the horizon with the in- and outgoing energy-momentum
is governed by the equation of motion \cite{thooft2,us}
\be
(\Delta - 1)\,\X^-(\theta,\phi) =
\int^{{}^{X^+}}\!\!\! dx^+\, T_{++}
(\theta,\phi)
\ee
and similar for $\X^+(\theta,\phi)$, where $\Delta$ denotes the angular
Laplacian. This equation of motion for the horizon is the direct analogue
of equation (\ref{nonu}) determining dynamical boundary, and it actually
gives an indication how one could include the angular coordinates
in our model.

A further correspondence between our approach and that of \cite{thooft2}
is that in both cases the fields $\X^\pm$ are canonically conjugate variables,
and the $S$-matrix is essentially the canonical
transformation that interchanges the role of $X^+$ and $X^-$.
In our theory, however, the variables $\X^\pm$ do not
describe the position of a permeable horizon, but the trajectory of a
{\it reflecting} boundary near $r=0$.
In this way we naturally obtain an $S$-matrix that is defined on the full
second quantized Hilbert space of the matter particles, while by analogy
with the moving mirror problem we can also explain the origin of Hawking
radiation.

\medskip

\medskip

\medskip

\noindent
{\bf Acknowledgements}

\noindent
We would like to thank T. Banks, C. Callan, S. Giddings,
A. Jevicki, G. Moore, K. Schoutens, S. Shatashvili, A. Strominger, L.
Susskind, L. Thorlacius,  and E. Witten for helpful
discussions. The research of H.V. is supported by NSF Grant PHY90-21984.
and that of E.V. by an Alfred P. Sloan Fellowship, the W.M. Keck
Foundation, the Ambrose Monell Foundation, and NSF Grant PHY 91-06210.

\noindent
{\renewcommand{\Large}{\normalsize}

\end{document}